\documentclass[11pt]{article}
\pdfoutput=1
\usepackage{epsfig,sint,cite,amsfonts,amssymb,amsmath,bbold}
\usepackage[font=small,labelfont={sf,bf}]{caption}
\usepackage{subcaption}
\usepackage[T1]{fontenc}
\usepackage{tikz}
\usetikzlibrary{decorations.pathreplacing}
\usetikzlibrary{patterns} %per usare i retini
\usepackage[english]{babel}
\usepackage{simplewick}
\usepackage{dsfont}
\usepackage[utf8]{inputenc}
\usepackage{physics}
\usepackage{siunitx}
\usepackage{booktabs}
\usepackage{wrapfig}
\usepackage{slashed}

\newcommand{\be}{\begin{equation}}
\newcommand{\ee}{\end{equation}}
\newcommand{\bea}{\begin{eqnarray}}
\newcommand{\eea}{\end{eqnarray}}
\newcommand{\bes}{\begin{eqnarray}}
\newcommand{\ees}{\end{eqnarray}}
\newcommand{\ba}{\begin{array}}
\newcommand{\ea}{\end{array}}
\newcommand{\Prj}{{\rm I}\!{\rm P}}
\def\dirac#1{\gamma_{#1}}
\def\diracstar#1#2{
    \setbox0=\hbox{$\gamma$}\setbox1=\hbox{$\gamma_{#1}$}
    \gamma_{#1}\kern-\wd1\kern\wd0
    \smash{\raise4.5pt\hbox{$\scriptstyle#2$}}}

\usetikzlibrary{decorations.markings}
\usetikzlibrary{svg.path}
\usetikzlibrary{shapes,fit}

\newcommand{\id}{\mathbb{1}}

\renewcommand{\i}{\mathrm{i}}

\begin{document}

\begin{titlepage}
%\begin{flushright}
%\hfill PREPRINT NUMBER \\
%\end{flushright}

$\;$ \\
\vspace{1.0cm}

\begin{center}
  
 {\Large\bf Four-dimensional factorization of the fermion determinant in lattice QCD\\[0.5ex]} 

\end{center}
\vskip 0.75 cm
\begin{center}
{\large  Leonardo Giusti and Matteo Saccardi}
\vskip 1.0cm
Dipartimento di Fisica, Universit\`a di Milano--Bicocca,\\
and INFN, sezione di Milano--Bicocca,\\
Piazza della Scienza 3, I-20126 Milano, Italy\\
\vskip 1.5cm

{\bf Abstract}
\vskip 0.35ex
\end{center}

\noindent
In the last few years it has been proposed a one-dimensional factorization of the fermion
determinant in lattice QCD with Wilson-type fermions that leads to a block-local action of the
auxiliary bosonic fields. Here we propose a four-dimensional generalization of this factorization.
Possible applications are more efficient parallelizations of Monte Carlo algorithms and codes,
master field simulations, and multi-level integration.
\vfill
\eject

\end{titlepage}

\section{Introduction\label{sec:intro}}
In path integrals of lattice gauge theories with fermions, once the Grassmann variables have been
analytically integrated out, the manifest locality of the action and of the observables is lost.
The fermion determinant is a non-local functional of the background gauge field, and the resulting
effective gauge theory is simulated with variants of the Hybrid Monte Carlo (HMC) algorithm~\cite{Duane:1987de}.
In the vast majority of cases, the algorithm implements global updates for an importance sampling
with a non-local action.

A few years ago it has been proposed a factorization of the gauge-field dependence of the fermion determinant
in lattice QCD based on a domain decomposition of the lattice~\cite{Luscher:2003qa,Luscher:2004pav,Ce:2016idq,Ce:2016ajy}.
The factorization has been derived in full details by decomposing the lattice in overlapping domains
along one of the dimensions only~\cite{Ce:2016ajy}. Once combined with the multi-boson idea~\cite{Luscher:1993xx}, it
leads to a local action in the block gauge, pseudofermion and multi-boson auxiliary fields~\cite{Ce:2016ajy}.
Extensive numerical tests have been performed since then~\cite{Ce:2016idq,Ce:2016ajy,Giusti:2017ksp,Ce:2017ndt}, and a
first computation of the hadronic vacuum polarization contribution to the anomalous magnetic moment of the muon
based on these ideas has been presented~\cite{DallaBrida:2020cik}.

The aim of this letter is to generalize the factorization of the fermion determinant
in Ref.~\cite{Ce:2016ajy} to four dimensions. This is not straightforward because, in a
multi-dimensional decomposition, the domains may not be naturally the union of disconnected regions.
The problem is solved by choosing judiciously a four-dimensional overlapping domain decomposition of the
lattice which leads to a simple block decomposition of the Dirac operator with highly-suppressed
elements in the off-diagonal blocks. These contributions can then be taken into account by introducing
multi-boson auxiliary fields.

A four-dimensional factorization of the gauge-field dependence of the fermion determinant boosts
our ability of simulating gauge theories with fermions, possibly triggering new perspectives in this field.
It allows for highly efficient parallelizations, also on heterogeneous architectures, of Monte Carlo
algorithms and of the corresponding codes by reducing very significantly the rate of data exchange
among different (blocks of) computer nodes where the various domains of the lattice are mapped to.  
In master field simulations~\cite{Luscher:2017cjh,Giusti:2018cmp,Francis:2019muy},  it allows for
a block-local accept/reject step in the HMC, solving the problem of the increasing numerical
precision needed for larger and larger volumes. Finally, a block-local action of the auxiliary bosonic
fields indeed opens the way to multi-level simulations of QCD in all four dimensions. 

The letter is organized as follows: in Section~\ref{sec:DD} we introduce the four-dimensional
domain decomposition of the lattice that we adopt, and in the following two Sections we derive the
factorization of the gauge-field dependence of the determinant. In Section~\ref{sec:MB} the residual interactions
among the various domains is taken exactly into account by introducing multiboson fields on their boundaries,
while in Section~\ref{sec:upd} a fully block-local Monte Carlo updating scheme is discussed. 
We end the letter with our conclusions and outlook. Notations, conventions, and technical details are
reported in several appendices.

\section{Four-dimensional domain decomposition of the lattice\label{sec:DD}}
\begin{figure}[t!]
\begin{center}
\includegraphics[width=0.8\columnwidth]{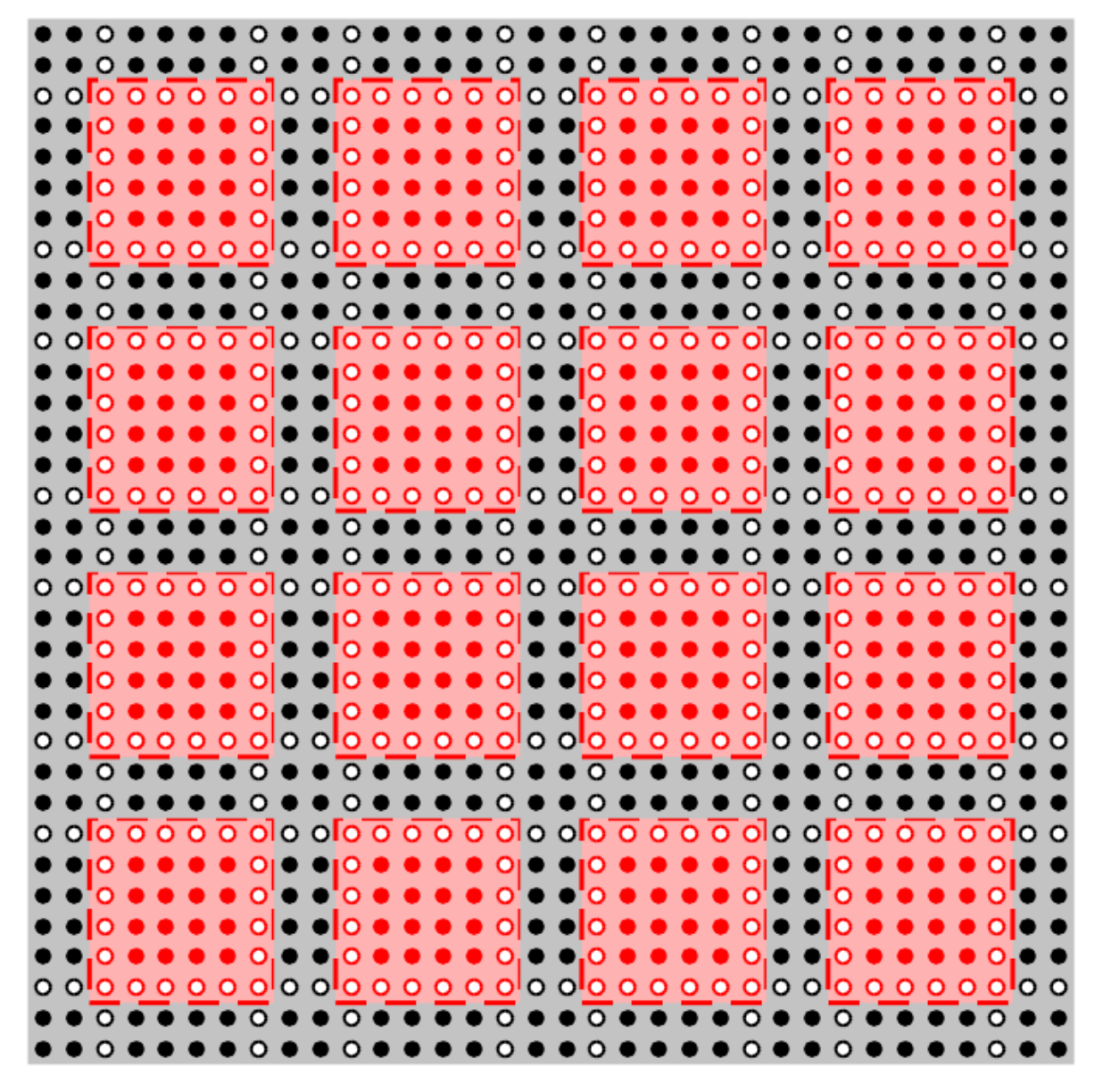}
\caption{Two-dimensional representation of the basic domain decomposition of the lattice in the disconnected domain
  $\Lambda_{0}$ (red square blocks) and the globally connected one $\Lambda_{1}$ (grey thick frame).
  The empty circles indicate the domain of hyperplanes $\partial \Pi$, see main text, with the red and black circles
  indicating $\partial\Lambda_{0}$ and $\partial \Pi_1$ respectively.
  \label{Fig:startingDD}}
\end{center}
\end{figure}
We consider a four-dimensional hyperrectangular lattice of spacing $a$ and lengths $L_\mu$ in the directions
labeled by $\mu=0,\dots,3$. We are interested in decomposing this lattice in all the four dimensions
by generalizing the one-dimensional domain decomposition introduced in Ref.~\cite{Ce:2016ajy}, see
also \cite{Giusti:2017ksp,Ce:2017ndt,DallaBrida:2020cik}. To this aim, we use some of the notation adopted in these papers by
assuming familiarity of the reader with them.
\begin{figure}[t!]
\begin{center}
\includegraphics[width=0.33\columnwidth]{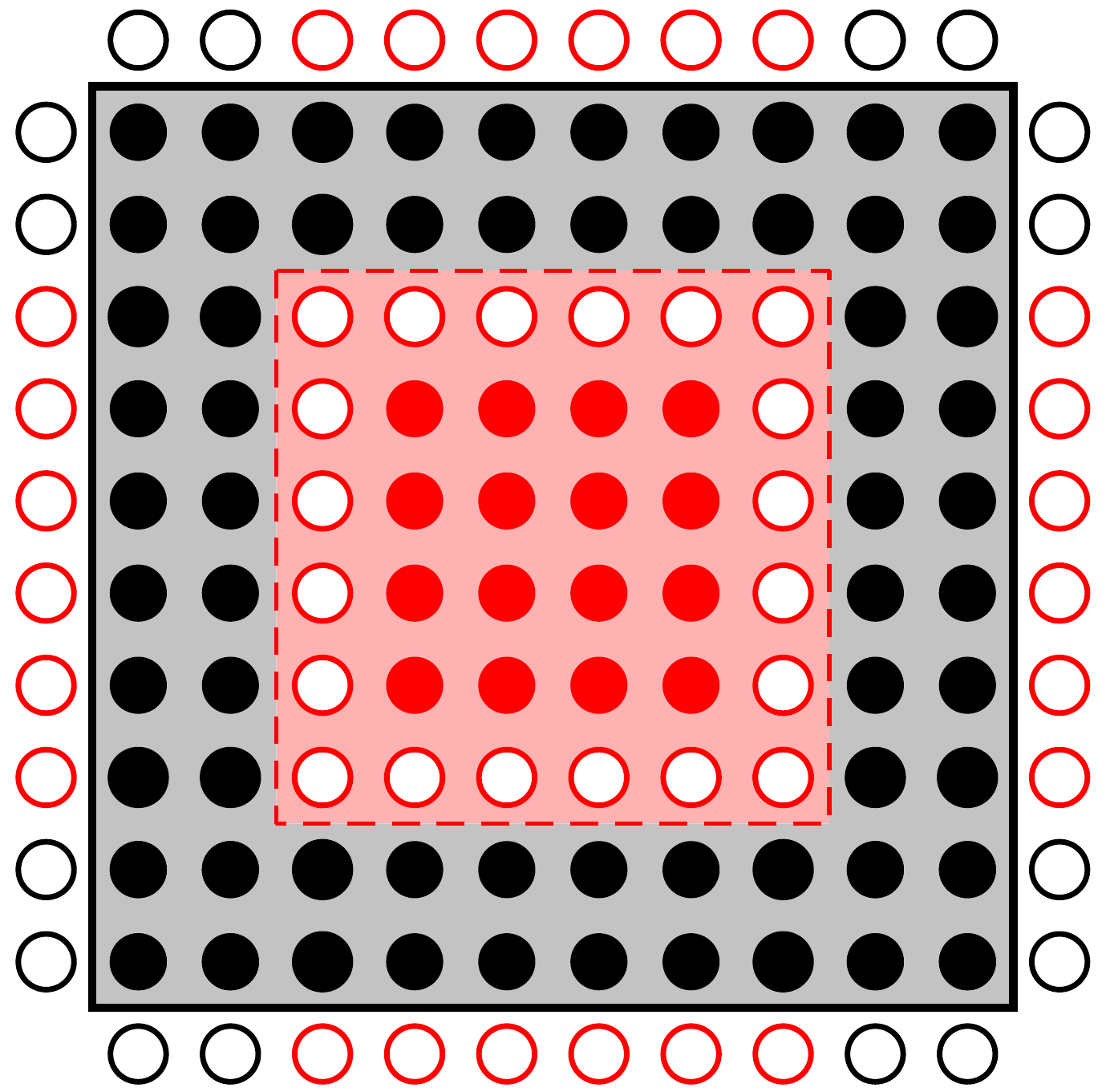}
\caption{A ``framed domain'' $\Omega^{\hat a}_0$ made by the union of a hyperrectangle $\Lambda_{0}^{\hat a}$ (red)
  and its frame $\Phi^{\hat a}_1$ (grey). The points of its exterior boundary $\partial \Omega^{\hat a *}_0$ are
  indicated with open circles outside the continuous black line. These circles are red, if they belong to
  $\partial \Lambda_{0}$, or black , if they belong to  $\partial\Pi_1$ and in particular to
  $\partial \bar\Omega^{\hat a *}_0$.\label{Fig:Omega0a}}
\end{center}
\end{figure}

We start by dividing the lattice in a domain $\Lambda_0$ made of hyperrectangular blocks embedded in a thick 
frame $\Lambda_1$. In the two-dimensional representation shown in Fig.~\ref{Fig:startingDD}, the 
blocks are represented by red squares, while the grey region is the frame. By construction $\Lambda_0$ is a
disconnected domain which can be decomposed as 
\be\label{eq:Lambda0}
\Lambda_0 = \bigcup_{\hat a} \Lambda^{\hat a}_0
\ee
where the label $\hat a$ identifies the single hyperrectangle, see Appendix~\ref{app:DD4d} for its definition.
The domain $\Lambda_1$ spans the entire lattice and it is connected\footnote{It is possible to introduce an even-odd decomposition
of the domain $\Lambda_{0}$, so that
the union of the even and the odd blocks plays the same r\^ole as the domains $\Lambda_{0}$ and
$\Lambda_{2}$ in the one-dimensional decomposition in Refs.~\cite{Ce:2016ajy,Giusti:2017ksp,Ce:2017ndt}.
The frame $\Lambda_{1}$ corresponds to the homologous one in these references.}, at variance of the
one-dimensional decomposition \cite{Ce:2016ajy}.
Typically the linear extension $B_\mu$ of the blocks in each direction
$\mu$ can be of a few fermi, while the thicknesses $b_\mu$ of the frame are typically of $0.5$~fm or so.
Following Refs.~\cite{Luscher:2003qa,Ce:2016ajy}, for each block
$\Lambda^{\hat a}_0$ we define 
\be
\partial \Lambda^{\hat a}_0\;,\quad \mbox{\rm and } \quad \bar \Lambda_0^{\hat a} = \Lambda_0^{\hat a}\backslash \partial \Lambda_0^{\hat a}\;,
\ee
where $\partial \Lambda^{\hat a}_0$ is the inner boundary of the block (open red circles in Fig.~\ref{Fig:startingDD})
defined as the set of points in $\Lambda^{\hat a}_0$ at a distance $a$ from the closest points of the lattice outside the block,
the latter being the exterior boundary $\partial \Lambda^{\hat a *}_0$. The sub-block $\bar \Lambda_0^{\hat a}$ is therefore the set of the inner
points of $\Lambda^{\hat a}_0$ (closed red circles in the same Figure). Analogously to Eq.~(\ref{eq:Lambda0}), it is useful
to define
\be
\partial \Lambda_0 = \bigcup_{\hat a} \partial \Lambda^{\hat a}_0\; , \qquad
\bar \Lambda_0 =  \bigcup_{\hat a} \bar\Lambda_0^{\hat a}\; .
\ee
The various boundary faces that form $\partial \Lambda_0$ belong to hyperplanes with normal
directions parallel to the axes of the lattice (open circles in Fig.~\ref{Fig:startingDD}). The planes
are spaced alternatively by ${B}_\mu$ and ${b}_\mu$ along each direction $\mu$, and their ensemble
is defined to be the domain $\partial \Pi$. The latter can be decomposed as 
\be
\partial \Pi = \partial \Lambda_0 \cup \partial \Pi_1\; ,    
\ee
where $\partial \Pi_1$ is represented by black open circles in Fig.~\ref{Fig:startingDD}. Notice
that $\partial \Pi_1$ belongs to $\Lambda_1$ and that
\be
\bar \Lambda_1 =  \Lambda_1 \backslash \partial \Pi_1\;  
\ee
is a disconnected domain. Each block $\Lambda^{\hat a }_0$ has an associated ``frame'' $\Phi^{\hat a}_1$ defined as the grey region
surrounding it, see Fig.~\ref{Fig:Omega0a} for a graphic representation and Appendix \ref{app:DD4d} for
its precise definition. The set of blocks $\Phi^{\hat a}_1$ clearly forms an overlapping domain
decomposition of $\Lambda_1$. The ``framed'' counterpart of $\Lambda^{\hat a }_0$ is given by  
\be
\Omega^{\hat a}_0 = \Lambda_0^{\hat a} \cup \Phi^{\hat a}_1\; ,
\ee
a definition which requires obvious modifications for the blocks
near the boundaries of the lattice, depending on the boundary conditions adopted.
The blocks $\Omega^{\hat a}_0$ form an overlapping
domain decomposition of the entire lattice $L$, see Fig.~\ref{Fig:startingDD}, similarly to what happens in
the one-dimensional case~\cite{Ce:2016ajy,Giusti:2017ksp}. Finally, we define
\be
\partial \Omega^{\hat a *}_0\; , \quad {\rm and} \quad
\partial \bar\Omega^{\hat a *}_0 = \partial \Omega^{\hat a *}_0 \cap \partial \Pi_1\; 
\ee
where $\partial \Omega^{\hat a *}_0$ is the exterior boundary of $\Omega^{\hat a}_0$, see
Fig.~\ref{Fig:Omega0a} for a graphic representation and Appendix~\ref{app:DD4d} for the definition,
while $\partial \bar\Omega^{\hat a *}_0$ is its subdomain belonging to $\partial \Pi_1$ (black open circles in
the same Figure).

In the next Sections we will need the projection operators to the subspace of quark fields supported on the
various sub-lattices, see Appendix~\ref{app:proj} for their definitions. We will indicate them with
the symbol $\Prj$ associated to a subscript indicating the sub-lattice considered, e.g.
$\Prj_{\Lambda^{\hat a }_0}$ for the block $\Lambda^{\hat a }_0$.

\section{Block decomposition of the fermion determinant\label{sec:detD}}
We are interested in factorizing the gauge-field dependence of the determinant of the
Wilson--Dirac operator $D$ defined in Eq.~(\ref{eq:Dwimp}) of Appendix~\ref{app:Dw}. To this aim, 
we start by decomposing the lattice $L$ as
\be
L = \partial\Lambda_{0} \cup \big[ \bar\Lambda_{0} \cup \Lambda_{1} \big]\; , 
\ee
and, accordingly, we rewrite $D$ as a $2 \times 2$ block matrix. By using Eq.~(\ref{eq:detblock}) in
Appendix~\ref{app:Schur}, the determinant can then be written as
\be\label{eq:rel1}
\det D = \det D_{\bar\Lambda_{0}} \det D_{\Lambda_{1}} \det \tilde D_{\partial\Lambda_{0}}\; ,
\ee
where\footnote{It is interesting to notice that $\tilde D_{\partial\Lambda_{0}}$ corresponds to the
effective Wilson--Dirac operator, once the Grassmann field variables
in $\bar\Lambda_{0}$ and $\Lambda_{1}$ have been
integrated out in the path integral. Analogous considerations apply to other Schur complements throughout the paper.} 
\be\label{eq:Dtilde0}
\tilde D_{\partial\Lambda_{0}} = {\bar D}_{\partial\Lambda_{0}} - D_{\partial\Lambda_{0},\Lambda_{1}} D_{\Lambda_{1}}^{-1}
D_{\Lambda_{1},\partial\Lambda_{0}}
\ee
and
\be\label{eq:Dbar0}
{\bar D}_{\partial\Lambda_{0}} = D_{\partial\Lambda_{0}} - D_{\partial\Lambda_{0},\bar\Lambda_{0}}
D_{\bar\Lambda_{0}}^{-1} D_{\bar\Lambda_{0},\partial\Lambda_{0}} = \sum_{\hat a}
{\bar D}_{\partial\Lambda_{0}^{\hat a}}\; .
\ee
In the formulas above and throughout the paper, the subscript of an operator indicates the domain where the
operator is restricted, e.g. $D_{\Lambda^{\hat a}_{0}}$ is the Wilson--Dirac operator restricted to the
domain $\Lambda^{\hat a}_{0}$ with Dirichlet boundary conditions imposed on its external boundaries. When the
subscript of the operator has two domains separated by a comma, this indicates a hopping term among these two domains,
see for instance Appendix~\ref{app:proj}. By noticing that
\be
D_{\partial\Lambda_{0},\Lambda_{1}} =  \sum_{\hat a} D_{\partial\Lambda^{\hat a}_{0},\Phi^{\hat a}_{1}}\;, \qquad
D_{\Lambda_{1},\partial\Lambda_{0}} = \sum_{\hat a} D_{\Phi^{\hat a}_{1},\partial\Lambda^{\hat a}_{0}}\; ,\\
\ee
it is clear that
\be\label{eq:Dextra}
D_{\partial\Lambda_{0},\Lambda_{1}} D_{\Lambda_{1}}^{-1} D_{\Lambda_{1},\partial\Lambda_{0}} =
\sum_{\hat a, \hat a'}
D_{\partial\Lambda^{\hat a}_{0},\Phi^{\hat a}_{1}}\, D_{\Lambda_{1}}^{-1}\, D_{\Phi^{\hat a'}_{1},\partial\Lambda^{\hat a'}_{0}}\; . 
\ee
If we decompose $\Lambda_1$ as the union of $\Phi^{\hat a}_{1}$ and its complement, the corresponding 
Schur decomposition of $D_{\Lambda_{1}}$, written in the $2 \times 2$ blocked form, allows us to rewrite
its inverse as in Eq.~(\ref{eq:Scmpt2}). This in turn implies that   
\bea
\Prj_{\Phi^{\hat a}_{1}} D_{\Lambda_{1}}^{-1} & = & D_{\Phi^{\hat a}_{1}}^{-1} - D_{\Phi^{\hat a}_{1}}^{-1}
D_{\Phi^{\hat a}_{1},\partial \bar\Omega^{\hat a *}_0} D_{\Lambda_{1}}^{-1}\; ,\label{eq:Schur1}\\[0.25cm]
D_{\Lambda_{1}}^{-1} \Prj_{\Phi^{\hat a}_{1}} & = & D_{\Phi^{\hat a}_{1}}^{-1} - D_{\Lambda_{1}}^{-1}
D_{\partial \bar\Omega^{\hat a *}_0,\Phi^{\hat a}_{1}} D_{\Phi^{\hat a}_{1}}^{-1}\; .\label{eq:Schur2}
\eea
By inserting Eqs.~(\ref{eq:Schur1}) and (\ref{eq:Schur2}) in Eq.~(\ref{eq:Dextra}), we obtain 
%\be\label{eq:DtildeII}
%\tilde D_{\partial\Lambda_{0}} = \hat D_{\partial\Lambda_{0}}  - \sum_{\hat a, \hat a'} 
%D_{\partial\Lambda^{\hat a}_{0},\Phi^{\hat a}_{1}} D_{\Phi^{\hat a}_{1}}^{-1} D_{\Phi^{\hat a}_{1},\partial \bar\Omega^{\hat a *}_0}
%D_{\Lambda_{1}}^{-1}
%D_{\partial \bar\Omega^{\hat a' *}_0,\Phi^{\hat a'}_{1}} D_{\Phi^{\hat a'}_{1}}^{-1} D_{\Phi^{\hat a'}_{1},\partial\Lambda^{\hat a'}_{0}}
%\ee
\be\label{eq:DtildeII}
\tilde D_{\partial\Lambda_{0}} = \hat D_{\partial\Lambda_{0}}  -
\hat D_{\partial\Lambda_{0},\partial \Pi_1} D_{\Lambda_{1}}^{-1} \hat D_{\partial \Pi_1,\partial\Lambda_{0}}
\ee
where
\bea
\hat D_{\partial\Lambda_{0},\partial \Pi_1} & = & - \sum_{\hat a} D_{\partial\Lambda^{\hat a}_{0},\Phi^{\hat a}_{1}}\, D_{\Phi^{\hat a}_{1}}^{-1}\,
D_{\Phi^{\hat a}_{1},\partial \bar\Omega^{\hat a *}_0}\; ,\\
\hat D_{\partial \Pi_1,\partial\Lambda_{0}} & = & - \sum_{\hat a} D_{\partial \bar\Omega^{\hat a *}_0,\Phi^{\hat a}_{1}} \,
D_{\Phi^{\hat a}_{1}}^{-1}\,
D_{\Phi^{\hat a}_{1},\partial\Lambda^{\hat a}_{0}}\; , 
\eea
and 
\be
\hat D_{\partial\Lambda_{0}} = \hat D^{\rm d}_{\partial\Lambda_{0}} + \hat D^{\rm h}_{\partial\Lambda_{0}}
\ee
with 
\be\label{eq:Dhat}
\hat D^{\rm d}_{\partial\Lambda_{0}} =  \sum_{\hat a} {\hat D}_{\partial\Lambda_{0}^{\hat a}}\,, \qquad
\hat D^{\rm h}_{\partial\Lambda_{0}} = \sum_{\hat a \neq \hat a'} \hat D_{\partial\Lambda_{0}^{\hat a},\partial\Lambda_{0}^{\hat a'}}\;,
\ee
and 
\bea\label{eq:Dhat_offdiag1}
{\hat D}_{\partial\Lambda^{\hat a}_{0}} & = & {\bar D}_{\partial\Lambda^{\hat a}_{0}} -
D_{\partial\Lambda^{\hat a}_{0},\Phi^{\hat a}_{1}}\, D_{\Phi^{\hat a}_{1}}^{-1}\, D_{\Phi^{\hat a}_{1},\partial\Lambda^{\hat a}_{0}}\;,\\[0.125cm] 
\hat D_{\partial\Lambda_{0}^{\hat a},\partial\Lambda_{0}^{\hat a'}} & = & - \frac{1}{2}\, D_{\partial\Lambda^{\hat a}_{0},\Phi^{\hat a}_{1}}
\Big[
D_{\Phi^{\hat a}_{1}}^{-1}  -
D_{\Phi^{\hat a}_{1}}^{-1}\, D_{\Phi^{\hat a}_{1},\partial \bar\Omega^{\hat a *}_0} D_{\Phi^{\hat a'}_{1}}^{-1} +\nonumber\\
& &\hspace{2.125cm} D_{\Phi^{\hat a'}_{1}}^{-1} -
D_{\Phi^{\hat a}_{1}}^{-1}\, D_{\partial \bar\Omega^{\hat a' *}_0,\Phi^{\hat a'}_{1}} D_{\Phi^{\hat a'}_{1}}^{-1}
\Big] D_{\Phi^{\hat a'}_{1},\partial\Lambda^{\hat a'}_{0}}\; .
\label{eq:Dhat_offdiag2}
\eea
Before proceeding further, it is already interesting to notice that
${\hat D}_{\partial\Lambda_{0}^{\hat a}}$ is the Schur complement of
$D_{\Omega_{0}^{\hat a}}$ with respect to the decomposition
$\Omega_{0}^{\hat a} = \partial\Lambda_{0}^{\hat a} \cup [\bar\Lambda_{0}^{\hat a} \cup \Phi^{\hat a}_{1}]$,
and that the hopping terms among the blocks
$D_{\partial\Lambda_{0}^{\hat a},\partial\Lambda_{0}^{\hat a'}}$ are suppressed with the thicknesses of the frame.
To manipulate the last sum on the r.h.s. of Eq.~(\ref{eq:DtildeII}), it is useful to define the Schur complement
\be\label{eq:DhatPi}
\hat D_{\partial \Pi_1} = D_{\partial \Pi_1} - D_{\partial \Pi_1,\bar \Lambda_1} D^{-1}_{\bar \Lambda_1} D_{\bar \Lambda_1,\partial \Pi_1}\; .  
\ee
Since $\partial \bar\Omega^{\hat a *}_0\in \partial\Pi_1$, in Eq.~(\ref{eq:DtildeII}) we can replace $D_{\Lambda_{1}}^{-1}$ with
its projection on $\partial \Pi_1$, which in turn is equal to $\hat D^{-1}_{\partial \Pi_1}$. Therefore, if we define the block matrix 
\be\label{eq:What}
\hat W =
\begin{pmatrix} 
\hat D_{\partial\Lambda_{0}} &  \hat D_{\partial\Lambda_{0},\partial \Pi_1} \\[0.25cm]
\hat D_{\partial \Pi_1,\partial\Lambda_{0}} &  \hat D_{\partial \Pi_1}\\
\end{pmatrix}
\; , 
\ee
it is immediate to see that
\be
\det \hat W = \det \hat D_{\partial \Pi_1} \det \tilde D_{\partial\Lambda_{0}} \;.
\ee
By remembering that
\be
\det D_{\Lambda_{1}} = \det D_{\bar \Lambda_{1}}  \det \hat D_{\partial \Pi_1}\; ,
\ee
Eq.~(\ref{eq:rel1}) can thus be written as
\be\label{eq:detD}
\det D = \det D_{\bar\Lambda_{0}} \det D_{\bar \Lambda_{1}} \det \hat W\;  .
\ee
Notice that the matrix $\hat W$ acts on the fermion fields defined on the domain
of the hyperplanes $\partial \Pi$  only. The off-diagonal blocks of
$\hat W$ are suppressed with the thicknesses of the frame of the blocks and 
depend on the gauge field in $\Lambda_1$ only. The one-dimensional decomposition
in Ref.~\cite{Ce:2016ajy} is readily obtained as a particular case of
Eq.~(\ref{eq:detD}) by noticing that in that case $\hat D_{\partial\Lambda_{0}}$ is
identified with $\hat W$ since the other blocks are absent, and that
$\hat D_{\partial\Lambda_{0}^{\hat a},\partial\Lambda_{0}^{\hat a'}}$ takes contribution from the first
and the third terms in the parenthesis in Eq.~(\ref{eq:Dhat_offdiag2}) only.

\section{Preconditioning of $\hat W$\label{sec:detW}}
By taking inspiration from the one-dimensional example, we would like to precondition
$\hat W$ so as to remain with a matrix which deviates from the identity by off-diagonal blocks which are
suppressed with the thicknesses of the frame. To this aim we first notice that
each block of the diagonal part $\hat D^{\rm d}_{\partial\Lambda_{0}}$ in Eq.~(\ref{eq:Dhat}) 
%\be
%\hat D_{\partial\Lambda_{0}}\!\! =\!\! \begin{pmatrix}
%    \ddots & \ddots \\
%    \ddots & {\hat D}_{\partial\Lambda_{0}^{\hat a}} & \!\!\!\!\!  \hat D_{\partial\Lambda_{0}^{\hat a},\partial\Lambda_{0}^{\hat a'}} \\[0.25cm]
%    & \!\!\!\!\! \hat D_{\partial\Lambda_{0}^{\hat a'},\partial\Lambda_{0}^{\hat a}} & {\hat D}_{\partial\Lambda_{0}^{\hat a'}}  &
%      \!\!\!\!\! \hat D_{\partial\Lambda_{0}^{\hat a'},\partial\Lambda_{0}^{\hat a''}} \\[0.25cm]
%    & & \!\!\!\!\! \hat D_{\partial\Lambda_{0}^{\hat a''},\partial\Lambda_{0}^{\hat a'}} & {\hat D}_{\partial\Lambda_{0}^{\hat a''}}  & \ddots \\[0.25cm]
%    & & & \ddots & \ddots
%  \end{pmatrix}\, ,\!\!\!\!\!
%\ee
depends on the gauge field in that (framed) block, while the elements of the off-diagonal component $\hat D^{\rm h}_{\partial\Lambda_{0}}$ are
suppressed with the thicknesses of the frame and depend on the gauge field in $\Lambda_1$ only. At variance of the one-dimensional case, 
here $\hat D_{\partial\Lambda_{0}}$ is not the only operator that appears in $\hat W$. We have to consider additional block
matrices, e.g. $\hat D_{\partial \Pi_1}$, because the domain $\Lambda_1$ is not factorized. The operator $\hat D_{\partial \Pi_1}$ may also
be decomposed in blocks similarly to $\hat D_{\partial\Lambda_{0}}$. For the factorization strategy of this letter, however, this
decomposition is not necessary and we proceed by considering this operator as a unique global domain.

The structure of $\hat D_{\partial\Lambda_{0}}$ suggests that we can define a preconditioned operator
$\overline W_1$ so that 
%\be\label{eq:Wbar}
%\hat W =\!\! \begin{pmatrix}
%    \ddots & \ddots \\
%    \ddots & {\hat D}_{\partial\Lambda_{0}^{\hat a}} & \!\!\!\!\!  0 \\[0.25cm]
%    & \!\!\!\!\! 0 & {\hat D}_{\partial\Lambda_{0}^{\hat a'}}  & \!\!\!\!\! 0 \\[0.25cm]
%    & & \!\!\!\!\! 0 & {\hat D}_{\partial\Lambda_{0}^{\hat a''}}  & \ddots \\[0.25cm]
%    & & & \ddots & \hat D_{\partial \Pi_1}
%  \end{pmatrix}
%\cdot 
%W_1\; ,
%\ee
\be\label{eq:Wbar}
\hat W = \left(
\begin{array}{c c}
\hat D^{\rm d}_{\partial\Lambda_{0}}  & 0              \\
               0               & \hat D_{\partial \Pi_1}
\end{array}
\right)\cdot \overline W_1\; ,
\ee
where 
%\be\label{eq:Wz}
%\hspace{-0.425cm} W_z \!\! =\!\!
%\left(
%\begin{array}{c c c c|c} 
%    \ddots & \ddots & & &\\
%    \ddots & z \Prj_{\partial\Lambda_{0}^{\hat a}} & \!\!\!\!\!  {\hat D}^{-1}_{\partial\Lambda_{0}^{\hat a}} \hat D_{\partial\Lambda_{0}^{\hat a},\partial\Lambda_{0}^{\hat a'}} & & \\[0.25cm]
%    & \!\!\!\!\!\!\!\!\!\! {\hat D}^{-1}_{\partial\Lambda_{0}^{\hat a'}} \hat D_{\partial\Lambda_{0}^{\hat a'},\partial\Lambda_{0}^{\hat a}} & z \Prj_{\partial\Lambda_{0}^{\hat a'}}  &
%      \!\!\!\!\!\!\! {\hat D}^{-1}_{\partial\Lambda_{0}^{\hat a'}} \hat D_{\partial\Lambda_{0}^{\hat a'},\partial\Lambda_{0}^{\hat a''}}\!\!\! &
%      \!\! W_{\partial\Lambda_{0},\partial \Pi_1}\!\!\! \\[0.25cm]
%    & & \!\!\!\!\! {\hat D}^{-1}_{\partial\Lambda_{0}^{\hat a'}} \hat D_{\partial\Lambda_{0}^{\hat a'},\partial\Lambda_{0}^{\hat a''}} & \ddots  &   \\[0.25cm]
%\hline
%& & & & \\
%& & W_{\partial \Pi_1,\partial\Lambda_{0}} & &\!\!\!\!\! z \Prj_{\partial\Pi_{1}} \\
%\end{array}
%\right)\,,
%\ee
\be\label{eq:Wz}
\hspace{-0.425cm} \overline W_z \!\! =\!\!
\left(
\begin{array}{c | c} 
z \Prj_{\partial\Lambda_{0}} + [{\hat D}^{\rm d}_{\partial\Lambda_{0}}]^{-1} \hat D^{\rm h}_{\partial\Lambda_{0}} &
\overline W_{\partial\Lambda_{0},\partial \Pi_1} \\[0.25cm]
\hline\\[-0.325cm]
\overline W_{\partial \Pi_1,\partial\Lambda_{0}} & z \Prj_{\partial\Pi_{1}}
\end{array}
\right)\,,
\ee
with $z\in\mathbb{C}$, 
\be
\overline W_{\partial\Lambda_{0},\partial \Pi_1} =  \sum_{\hat a} \Prj_{\partial\Lambda^{\hat a}_{0}} D_{\Omega^{\hat a}_{0}}^{-1}\, D_{\Phi^{\hat a}_{1},\partial \bar\Omega^{\hat a *}_0}\; ,
\ee
and
\be
\overline W_{\partial \Pi_1,\partial\Lambda_{0}} = \hat D_{\partial \Pi_1}^{-1} \hat D_{\partial \Pi_1,\partial\Lambda_{0}}\, .
\ee
Notice that the off-diagonal block operators of $\overline W_z$ act on a subspace of $\partial\Pi$ identified by the projector
$P_{\partial\Pi}=P_{\partial\Lambda_{0}}+P_{\partial\Pi_1}$ defined in Appendix~\ref{app:proj}. Indeed at variance of
$\Prj_{\partial\Lambda_{0}}$ and $\Prj_{\partial\Pi_1}$, the projectors $P_{\partial\Lambda_{0}}$ and
$P_{\partial\Pi_1}$ include also the appropriate projectors on the spinor index for the
inner and outer boundaries of the blocks $\Lambda^{\hat a}_{0}$ and $\Omega^{\hat a}_0$ respectively. As shown in
Eq.~(\ref{eq:reduced}) of Appendix~\ref{app:Schur}, it then holds
\be\label{eq:bella0}
\det \overline W_1 = \det W_1\, , \qquad {\rm where } \qquad W_z = P_{\partial\Pi}\, \overline W_z\, P_{\partial\Pi}\, ,
\ee
with the dimensionality of the matrix $W_z$ being smaller by essentially a factor 2 with respect to the one of $\overline W_z$.
By combining Eqs.~(\ref{eq:detD}), (\ref{eq:Wbar}) and (\ref{eq:bella0}) we obtain the final result
\be\displaystyle\label{eq:fact1}
\det D = \frac{1}{\displaystyle \det D^{-1}_{\Lambda_{1}} \prod_{\hat a}\left[ \det D_{\Phi^{\hat a}_{1}} \det  D_{\Omega^{\hat a}_{0}}^{-1}\right]}\, \det W_1\;  .
\ee
The denominator in Eq.~(\ref{eq:fact1}) has already a factorized dependence on the gauge field in the various blocks of $\Lambda_0$.
The next Section will be dedicated to the factorization of the remaining global contribution $\det W_1$.

\section{Multi-boson factorization of $\det W_1$\label{sec:MB}}
For large enough thicknesses of the frame $\Lambda_1$, we expect the matrix $W_1$ to have   
a large spectral gap, a fact which makes it effective to express its determinant through
a polynomial approximation of $W^{-1}_1$. As reviewed in the Appendix D of Ref.~\cite{Ce:2016ajy},
a generalization of L\"uscher's original multiboson proposal~\cite{Luscher:1993xx} to complex
matrices~\cite{Borici:1995np,Borici:1995bk,Jegerlehner:1995wb} starts by approximating the function $1/z$,
with $z\in\mathbb{C}$, by the polynomial
\be\label{eq:bella}
P_N(z) \equiv \frac{1-R_{N+1}(z)}{z} = c_N \prod_{k=1}^N (z-z_k) \;,
\ee
where $N$ is chosen to be even, the $N$ roots of $P_N(z)$ are obtained by requiring that
for the remainder polynomial $R_{N+1}$ it holds $R_{N+1}(0)=1$, and $c_N$ is an irrelevant numerical
constant. The roots $z_k$ can be chosen to lie on an ellipse passing through the origin of the
complex plane with center $1$ and foci
$1 \pm c$,
\begin{equation}
u_k = 1- z_k =\cos{\left(\frac{2\pi k}{N+1}\right)} + i \sqrt{1-c^2} \sin{\left(\frac{2\pi k}{N+1}\right)}\; ,
\quad k=1,\dots,N  \;.
\end{equation}
This polynomial can be used to approximate the inverse determinant as 
\be\label{eq:bella2}
\det \overline W_1\,  = \frac{\det\{1-R_{N+1}(\overline W_1)\}}{\det P_N(\overline W_1)}\; ,
\ee
where, if the moduli of all eigenvalues of $\overline W_1$ are smaller than $1$, the numerator of the
r.h.s. converges exponentially to $1$ as $N$ is increased. Thanks to the $\gamma_5$
hermiticity of $\hat W$ and of ${\hat D}^{\rm d}_{\partial\Lambda_{0}}$, the matrix $\overline W_1$ can be written
as a product of two Hermitian matrices which in turn implies that $\overline W_1$ is similar to
$\overline W^\dagger_1$. Since the $z_k$ come in complex conjugate pairs, the approximate determinant
can then be written in a manifestly positive form,
\begin{equation}
\frac{1}{\det\{ P_N(\overline W_1)\}} =  C \prod_{k=1}^{N/2} {\det}^{-1} \big\{ (z_k -\overline W_1 )^\dagger (z_k -\overline W_1) \big\} \\[0.25cm]
 = C \prod_{k=1}^{N/2} {\det}^{-1} \left[ \overline W_{u_k}^\dagger \, \overline W_{u_k}\right]
\label{e:pV}
\end{equation}
where $C$ is again an irrelevant numerical constant and $\overline W_z$ is defined in Eq.~(\ref{eq:Wz}). As a result
\be\displaystyle\label{eq:fact2}
\frac{\det D}{\det\{1-R_{N+1}(W_1)\}} \propto
\frac{1}{\displaystyle\det D^{-1}_{\Lambda_{1}} \prod_{\hat a}\left[ \det D_{\Phi^{\hat a}_{1}} \det  D_{\Omega^{\hat a}_{0}}^{-1}\right]
\prod_{k=1}^{N/2} {\det} \left[ W_{u_k}^\dagger \, W_{u_k}\right]}\, , 
\ee
where we have replaced $\overline W_{u_k}$ with $ W_{u_k}$ by using again the first relation in Eq.~(\ref{eq:bella0})
which, for $z\neq 1$,
is valid up to an irrelevant multiplicative
constant. The first factor and the first product in the denominator on the r.h.s.
can be included in the effective gluonic action
via standard pseudofermions defined within the blocks labeled by the subscript of the operators. 

\subsection{Multiboson action}
Each of the $N/2$ factors in the last product in the denominator of the r.h.s. of Eq.~(\ref{eq:fact2})
can be represented, up to an irrelevant multiplicative constant, as 
\be
\frac{1}{{\det} \left[W_{u_k}^\dagger \, W_{u_k}\right]} \propto
\int d\chi_k d\chi_k^\dagger\; e^{\displaystyle-|W_{u_k} \chi_k|^2 }\; . \\
\ee
The $N/2$ multiboson fields $\chi_k$ are defined on the subspace of $\partial\Pi$ identified by the projector
$P_{\partial\Pi}$. Each of them can be decomposed as $\chi=\chi_{_{\partial\Lambda_{0}}}+\chi_{_{\partial\Pi_1}}$, with
$\chi_{_{\partial\Lambda_{0}}}=P_{\partial\Lambda_{0}} \chi$ and $\chi_{_{\partial\Pi_1}}=P_{\partial\Pi_1} \chi$. As a result
\begin{equation}
\label{eq:multiboson}
\begin{split}
|W_{z} \chi|^2 &=
\sum_{\hat a} \Big|P_{\partial\Lambda_{0}^{\hat a}}\Big[z\,\chi_{_{\partial\Lambda_{0}}} +
[{\hat D}^{\rm d}_{\partial\Lambda_{0}^{\hat a}}]^{-1} \hat D^{\rm h}_{\partial\Lambda_{0}} \chi_{_{\partial\Lambda_{0}}}
+ D_{\Omega^{\hat a}_{0}}^{-1}\, D_{\Phi^{\hat a}_{1},\partial \bar\Omega^{\hat a *}_0} \chi_{_{\partial\Pi_1}}\Big] \Big|^2
\; \\
&+ \Big|z\, \chi_{_{\partial\Pi_1}} + W_{\partial \Pi_1,\partial\Lambda_{0}} \chi_{_{\partial\Lambda_{0}}} \Big|^2\; .
\end{split}
\end{equation}
The term on the second line of the r.h.s of Eq.~(\ref{eq:multiboson}) depends on the gauge field
in $\Lambda_1$ only. The gauge field within the domain $\Lambda_0$ appears only on the
first line. As a result {\it the 
dependence of the multi-boson action from the gauge field in the blocks $\Lambda^{\hat a}_0$
is factorized}. Moreover, all contributions in Eq.~(\ref{eq:multiboson}) are highly suppressed
with the thicknesses of the frame. This implies that the order $N$ of the multi-boson polynomial
can be rather low, i.e. of the order of ten or so~\cite{Ce:2016ajy}.

\subsection{Reweighting factor}
 A given correlation function of a string of fields $O$ can finally be written as 
\be
\langle O \rangle  = \frac{\langle O\, {\cal W}_N \rangle_N}{\langle {\cal W}_N \rangle_N}\; ,
\ee
where $\langle\cdot\rangle_N$ indicates the expectation value for an importance sampling 
with $N$ multi-bosons in the action, and 
\be
{\cal W}_N = \det\{1-R_{N+1}(W_1)\}\,. 
\ee
By using Eq.~(\ref{eq:bella}), up to an irrelevant numerical
multiplicative constant, the reweighting factor can be written as 
\be
{\cal W}_N \propto \frac{1}{\displaystyle {\det}\,\left\{ W^{-1}_1 
\prod_{k=1}^{N/2}  [W^{\dagger}_{u_k}]^{-1} W^{-1}_{u_k} \right\}}\; ,
\ee
a representation which suggests the random noise estimator 
\be\label{eq:WNeta}
{\cal W}_N = \frac{\int [d \eta] [d\eta^\dagger]e^{-\xi^\dagger W^{-1}_1 \xi}}{\int [d \eta] [d\eta^\dagger] e^{-\eta^\dagger \eta}}\; ,\qquad
\xi = \prod_{k=1}^{N/2}  W^{-1}_{u_k} \eta\; .
\ee
The expectation value can then be computed as
\bea\label{eq:rwgt}
\langle O \rangle  = \frac{\langle O\, {\cal W}_N \rangle_N}{\langle {\cal W}_N \rangle_N}
& = & \langle O_{\rm fact}\, \rangle_N +
\big\langle O\,  \frac{{\cal W}_N}{\langle {\cal W}_N \rangle_N} - O_{\rm fact} \big\rangle_N\; ,
\eea
where $O_{\rm fact}=O$ if the observable is already factorized, otherwise it has to be a rather precise factorized
approximation of $O$ (see Ref.~\cite{Ce:2016idq} for instance). As a result, 
$\langle O_{\rm fact}\, \rangle_N$ can be computed with a fully factorized integration algorithm, while the last (small)
contribution on the r.h.s. of Eq.~(\ref{eq:rwgt}) can be estimated in the standard way.

\section{Block-local updates\label{sec:upd}}
The factorization of the fermionic contribution to the effective gluonic action in
Eqs.~(\ref{eq:fact2})--(\ref{eq:multiboson}) allows for a decoupling of the link variables
in different blocks $\Lambda_{0}^{\hat a}$.  This can be achieved by generalizing the Domain
Decomposed Hybrid Monte Carlo (DD-HMC) proposed many years ago~\cite{Luscher:2004pav}
to a MultiBoson Domain Decomposed Hybrid Monte Carlo (MB-DD-HMC)~\cite{Ce:2016ajy}. To
this aim, the molecular dynamics evolution is restricted to the subset of all link variables,
referred to as the {\it active link variables}, which have both endpoints in the same block
$\Lambda_{0}^{\hat a}$ and at most one endpoint on the inner boundary of the block
(white open circles in Fig.~\ref{Fig:startingDD}). From Eqs.~(\ref{eq:fact2})--(\ref{eq:multiboson}),
it is clear that the active link variables in different blocks are decoupled from each other during
the molecular dynamics evolution because the multiboson fields and the inactive gauge links are
kept constant in this phase of the simulation. The accept/reject step
can thus be carried out independently on each block $\Lambda_{0}^{\hat a}$, i.e. there will be
blocks where the proposed new configuration is accepted and blocks where it is not.
In between every update cycle, the gauge field is then translated by a random vector $v$, i.e.
\be
U_\mu(x)\rightarrow U_\mu(x+v)\; , 
\ee
to ensure that all link variables are treated equally on average. Before restarting the
molecular dynamics evolution, new pseudofermion and multiboson fields need to be
generated. The pseudofermions can be generated locally in each block $\Lambda_{0}^{\hat a}$.
The multibosons, instead, require a global inversion of the Dirac operator but on a vector
belonging to the domain $\partial\Pi$ which is much smaller than the entire lattice\footnote{The localization
of the generation of the multiboson fields is beyond the scope of this paper.}.

In such an updating scheme, one needs to be sure that a good fraction of the link variables
can be updated in each step. This is the case if the linear extensions of the blocks are
at least of a few fermi. If, for instance, we consider blocks with an extension of $2.5$~fm and a frame 
of $0.5$~fm in all directions, the fraction of the active links is
readily computed to be approximatively $50\%$, a value which increases very rapidly
with the size of the blocks.

\section{Conclusions and outlook}
The factorization of the gauge-field dependence of the fermion determinant clearly
boosts our ability of simulating gauge theories in the presence of fermions. In particular the
complete four-dimensional factorization of the molecular dynamics evolution and of
the accept/reject steps may change the way we simulate lattice gauge theories
in several ways:\\ 

\noindent {\it Parallelization $-$} During the molecular dynamics evolution and in the accept-reject
step the link variables in different blocks are decoupled from each other, and the HMC runs independently in each block.
On heterogeneous architectures, one can envisage to simulate each block on a sub-set of nodes which have
faster connections (or, for instance, on a single GPU) without the need to communicate during long periods of
simulation time. A communication overhead is required only when the gauge field is shifted and the multiboson
fields are generated. This is typically a very small fraction of the computer time of the simulation.\\

\noindent {\it Master field simulations $-$} During the molecular dynamics evolution and 
for the accept-reject step, an inversion of the global lattice Dirac operator is never required. In master
field simulations in the presence of fermions~\cite{Luscher:2017cjh,Giusti:2018cmp,Francis:2019muy},
this solves the problem of the increasing numerical precision needed for inverting the Dirac operator on
larger and larger volumes.\\

\noindent {\it Multi-level integration $-$} The update procedure sketched in Section \ref{sec:upd} calls for a
two-level Monte Carlo integration scheme~\cite{Ce:2016ajy} where first $n_0$ {\it level-$0$} independent configurations of the
gauge field are generated over the entire lattice, and then for each of them $n_1$ {\it level-$1$} configurations
of the active links are generated by keeping fixed the inactive links and the multiboson fields. The two-level
estimate of an observable is then computed by averaging over the $n_0\cdot n^{n_{\rm b}}_1$ configurations obtained at
a cost proportional to $n_0\cdot n_1$, where $n_{\rm b}$ is the number of blocks in $\Lambda_0$. This two-level integration
can in principle be generalized to a multi-level scheme by iterating the domain decomposition and the integration
procedure. Extensive numerical tests which have been performed in the one-dimensional
case~\cite{Ce:2016idq,Ce:2016ajy,Giusti:2017ksp,Ce:2017ndt,DallaBrida:2020cik} have already shown the benefit of the
multi-level integration in solving the signal to noise ratio problem in the computation of correlation functions in
lattice QCD.

\section{Acknowledgments}
L.G. thanks Martin L\"uscher for many illuminating discussions over the
years on the topic of this letter.

\appendix

\section{$O(a)$-improved Wilson-Dirac operator\label{app:Dw}}
The massive $O(a)$-improved Wilson-Dirac operator is defined
as\footnote{Throughout this appendix the lattice spacing is set to unity for notational simplicity.}
\cite{Sheikholeslami:1985ij,Luscher:1996sc}
\be\label{eq:Dwimp}
D = D_{\rm w} + D_{\rm sw} + m_{0} \; ,
\ee
where $m_{0}$ is the bare quark mass, $D_{\rm w}$ is the massless Wilson-Dirac operator
\be\label{eq:Dw}
D_{\rm w} = \frac{1}{2} \{ \gamma_{\mu}(\nabla_{\mu}^{*} + \nabla_{\mu} ) - \nabla_{\mu}^{*} \nabla_{\mu} \} \; ,
\ee
with $\gamma_{\mu}$ being the Dirac matrices and the summation over repeated indices is understood.
The covariant forward and backward derivatives $\nabla_{\mu}$ and $\nabla_{\mu}^{*}$ are defined to be
\be\label{eq:cov}
\nabla_{\mu}\psi(x) = U_{\mu}(x)\psi(x+\hat{\mu})-\psi(x),\;\;\nabla_{\mu}^{*}\psi(x) = \psi(x) - U^{\dagger}_{\mu}(x-\hat{\mu})\psi(x-\hat{\mu})\;,
\ee
where $U_{\mu}(x)$ are the link fields and $\hat{\mu}$ is the unit versor along the direction $\mu$. By inserting
Eq.~(\ref{eq:cov}) in Eq.(\ref{eq:Dw}), the Wilson operator reads
\be\label{DwPsi}
\hspace{-0.05cm}D_{\rm w}\psi(x)\! =\! 4\psi(x) - \frac{1}{2}\! \sum_{\mu=0}^{3}\! \left\{ U_{\mu}(x)(1-\gamma_{\mu})\psi(x+\hat{\mu}) + U^{\dagger}_{\mu}(x-\hat{\mu})(1+\gamma_{\mu})\psi(x-\hat{\mu})\! \right\}\, .
\ee
The second term on the r.h.s. of Eq.~(\ref{eq:Dwimp}) is the Sheikholeslami-Wohlert operator defined as 
\be
D_{\rm sw}\psi(x) = c_{_{SW}} \frac{i}{4} \sigma_{\mu\nu} \widehat{F}_{\mu\nu}(x) \psi(x) \; ,
\ee
where $\sigma_{\mu\nu}=\frac{i}{2}[\dirac\mu,\dirac\nu]$, and $\widehat F_{\mu\nu}(x)$ is
the clover discretization of the field strength tensor which is given by
\be
\widehat{F}_{\mu\nu}(x) = \frac{1}{8} \{ Q_{\mu\nu}(x) - Q_{\nu\mu}(x) \}\; , 
\ee
with
\be
\begin{split}
Q_{\mu\nu}(x) &= U_{\mu}(x)\,U_{\nu}(x+\hat{\mu})\,U_{\mu}^{\dagger}(x+\hat{\nu})\,U_{\nu}^{\dagger}(x)
\\ & + U_{\nu}(x)\,U^\dagger_{\mu}(x-\hat{\mu}+\hat{\nu})\,U_{\nu}^{\dagger}(x-\hat{\mu})\,U_{\mu}(x-\hat{\mu})
\\ & + U_{\mu}^{\dagger}(x-\hat{\mu})\,U_{\nu}^{\dagger}(x-\hat{\mu}-\hat{\nu})\,U_{\mu}(x-\hat{\mu}-\hat{\nu})\,U_{\nu}(x-\hat{\nu})
\\ & + U_{\nu}^{\dagger}(x-\hat{\nu})\,U_{\mu}(x-\hat{\nu})\,U_{\nu}(x+\hat{\mu}-\hat{\nu})\,U_{\mu}^{\dagger}(x) \; .
\end{split}
\ee
It is also possible to use the alternative expression for $D_{\rm sw}$ given by
\be
D_{\rm sw} + (4+m_0) \rightarrow (4+m_{0}) \exp{\frac{c_{_{SW}}}{4+m_{0}} \frac{i}{4} \sigma_{\mu\nu} \widehat{F} _{\mu\nu} } \; ,
\ee
which has been proposed in the context of master field simulations~\cite{Francis:2019muy}.

\section{Definitions of basic domains\label{app:DD4d}}
\begin{figure}[t!]
\begin{center}
\includegraphics[width=0.33\columnwidth]{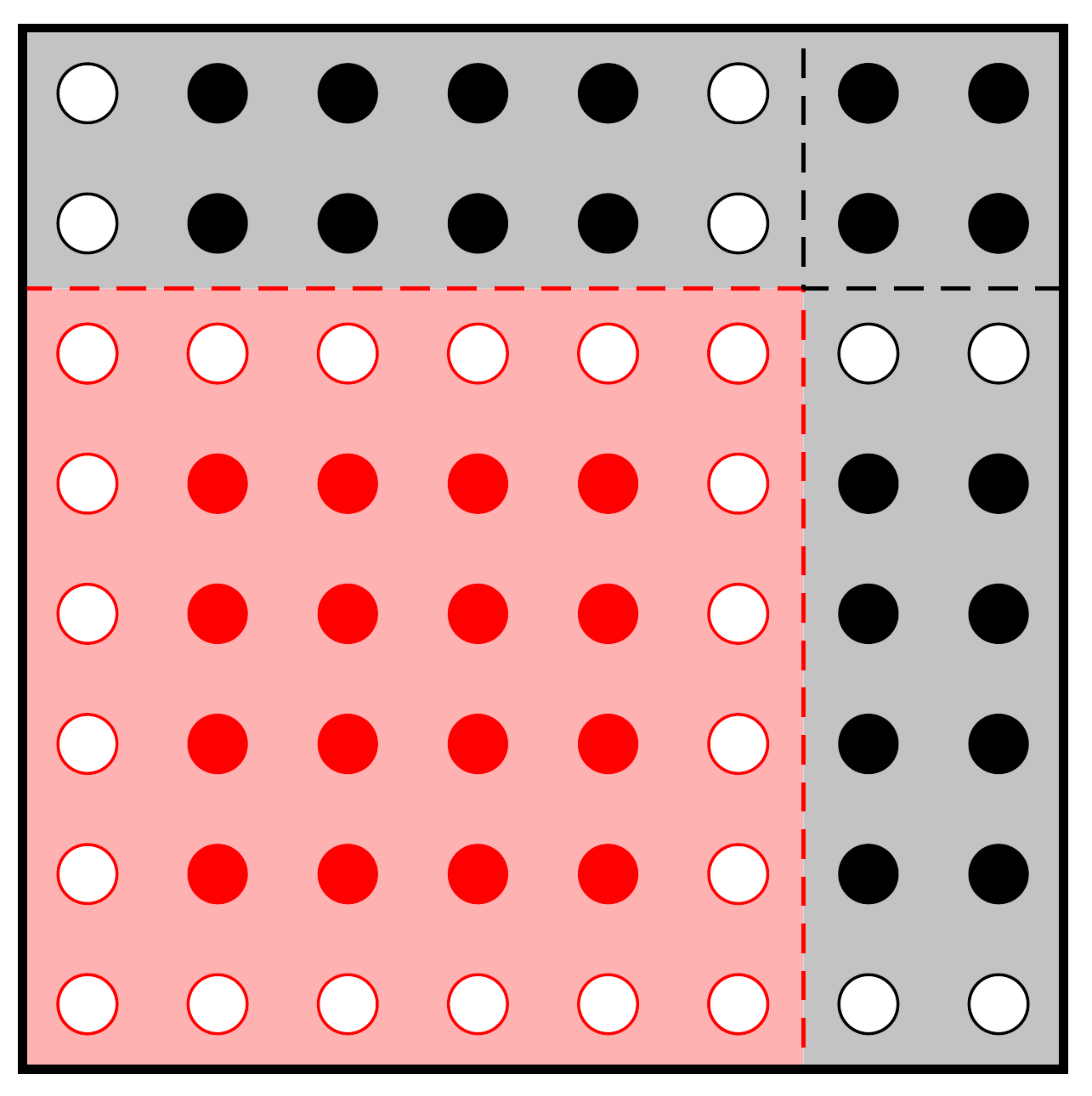}
\caption{Two-dimensional representation of a basic cell $\Gamma^{\hat a}$.\label{Fig:basic_cell}}
\end{center}
\end{figure}
To easily label the various subdomains considered in this letter, it is useful to introduce a non-overlapping domain decomposition of the
lattice so that the entire lattice $L$ is decomposed as 
\be
L = \bigcup_{\hat a} \Gamma^{\hat a}\; ,
\ee
where $\Gamma^{\hat a}$ is a basic hyperrectangular cell, see Fig.~\ref{Fig:basic_cell} for a 2-dimensional representation.
Each cell has dimension $G_\mu={B}_\mu + {b}_\mu$ in the direction $\mu$ and it is uniquely identified by the position of
its lower-left corner, given in four-dimensional Cartesian coordinates (in units of $G_{\mu}$) by $\hat a=\{a_0,a_1,a_2,a_3\}$,
where 
\be
a_\mu = 0,\dots, \frac{L_\mu}{G_\mu} -1\;, \qquad \mu=0,\dots,3\; ,
\ee
where $L_\mu$ is the length of the lattice along direction $\mu$. As a result, the global lattice coordinates of the lower-left point of the basic cell
are given by $x_\mu = G_\mu \cdot a_\mu$ (no summation over repeated indices is meant here).

To map the blocks of the decomposition in Fig.~\ref{Fig:startingDD} to the basic cells, the latter are further decomposed in
$2^4$ blocks as depicted in Fig.~\ref{Fig:basic_cell}. Within
each cell, the $16$ blocks can be identified by their local Cartesian coordinates in each direction $\mu$, i.e.
by $\hat d=(d_0,d_1,d_2,d_3)$ with $d_\mu=0,1$. In particular, the lower-left block ($\hat d = \hat 0$) of $\Gamma^{\hat a}$
identifies the block $\Lambda_{0}^{\hat a}$ of $\Lambda_{0}$, with their lower-left corners coinciding. The other blocks of
the basic cell belong to $\Lambda_1$, and the coordinates of their lower-left point are given by
$x_\mu = G_\mu \cdot a_\mu + B_\mu \cdot d_\mu$ with $\hat d \neq \hat 0$.
With those definitions we can finally write
\be
\Gamma^{\hat a} = \Lambda_{0}^{\hat a} \bigcup_{\substack{\hat d\neq \hat 0 \\ d_{\mu} = 0,1}} \Lambda_1^{(\hat a, \hat d)}\,  .
\ee
For each block $\Lambda^{\hat a }_0$, it is useful to define its ``frame'' $\Phi_{1}^{\hat a}$, which is shown in Fig.~\ref{Fig:Omega0a}, as
\be
\Phi_{1}^{\hat a} = \bigcup_{\substack{(\hat c, \hat d)\neq (\hat 0, \hat 0) \\[0.125cm] {c_\mu,d_{\mu} = 0,1} \big| (d-c)_\mu=0,1}}
\Lambda_1^{(\hat a -\hat c, \hat d)}\, .
\ee
Therefore, the ``framed'' domain
\be
\Omega_{0}^{\hat a} = \Lambda_{0}^{\hat a} \cup \Phi_{1}^{\hat a}\, 
\ee
is made of $3^4$ blocks with the obvious modifications for the blocks near the boundaries of the lattice depending on the boundary conditions adopted.
The blocks $\Phi^{\hat a}$ clearly form an overlapping domain decomposition of the entire domain $\Lambda_1$. Analogously, the blocks $\Omega^{\hat a}_0$
form an overlapping domain decomposition of the entire lattice $L$, similarly to what happens in the one-dimensional case~\cite{Ce:2016ajy,Giusti:2017ksp}.

\section{Projectors\label{app:proj}}
In this Appendix we define projectors on the various domains introduced in Section~\ref{sec:DD}.  
For $\Lambda_0^{\hat a}$ the projector is defined as 
\be\label{eq:proj_bulk}
\Prj_{\Lambda_{0}^{\hat a}} \psi(x) = \begin{cases}
\psi(x) & \text{if } x \in \Lambda_{0}^{\hat a}\, ,\\
0 & \text{otherwise}\, ,\\
\end{cases}
\ee
i.e. it localizes the quark field $\psi(x)$ inside the domain indicated in the subscript. It follows
that
\be
\Prj_{\Lambda_{0}} = \sum_{\hat a} \Prj_{\Lambda_{0}^{\hat a}}\; .
\ee
Projectors on other domains, e.g. $\Prj_{\Omega^{\hat a}_0}$, $\Prj_{\Omega_0}$, $\Prj_{\partial\Lambda_{0}^{\hat a}}$, $\Prj_{\partial\Pi}$, etc.,
are defined analogously.

Projectors on the inner and outer boundaries of $\Lambda_{0}^{\hat a}$ are indicated with
$P_{\partial\Lambda_{0}^{\hat a}}$ and $P_{\partial\Lambda_{0}^{\hat a *}}$ respectively, and they are defined so that 
\be
P_{\partial\Lambda_{0}^{\hat a}} D_{\partial\Lambda_{0}^{\hat a},\partial\Lambda_{0}^{\hat a *}} = D_{\partial\Lambda_{0}^{\hat a},\partial\Lambda_{0}^{\hat a *}}
P_{\partial\Lambda_{0}^{\hat a *}}\,,\qquad
P_{\partial\Lambda_{0}^{\hat a *}} D_{\partial\Lambda_{0}^{\hat a *},\partial\Lambda_{0}^{\hat a}} =
D_{\partial\Lambda_{0}^{\hat a *},\partial\Lambda_{0}^{\hat a}} P_{\partial\Lambda_{0}^{\hat a}}\;.
\ee
From Eq.~(\ref{DwPsi}) it holds
\be
\begin{gathered}
\big[D_{\partial\Lambda_{0}^{\hat a *},\partial\Lambda_{0}^{\hat a}} \psi\big](x) = -\Prj_{\partial\Lambda_0^{\hat a *}}
\sum_{\mu = 0}^{3} \left[U_{\mu}(x)\, \frac{1-\gamma_{\mu}}{2}\, \Prj_{\partial\Lambda_0^{\hat a}}\psi(x+\hat{\mu}) \right. \\
\left. + U_{\mu}^{\dagger}(x-\hat{\mu})\, \frac{1+\gamma_{\mu}}{2}\,\Prj_{\partial\Lambda_0^{\hat a}}\psi(x-\hat{\mu}) \right] \;,
\end{gathered}
\ee
and analogously for $D_{\partial\Lambda_{0}^{\hat a},\partial\Lambda_{0}^{\hat a *}}$ with
$\Prj_{\partial\Lambda_0^{\hat a}} \longleftrightarrow \Prj_{\partial\Lambda_0^{\hat a *}}$. This implies that
\be\label{eq:proj_inner}
P_{\partial\Lambda_{0}^{\hat a}}\,\psi(x) = \begin{cases}
   0 & \text{if } x \notin \partial\Lambda_{0}^{\hat a}\, , \\[0.125cm]
\displaystyle   \frac{1-\gamma_{\mu}}{2}\, \psi(x) & \text{if } x\in\partial\Lambda_{0}^{\hat a} \text{ and } \exists! \; \mu\,
\big|\; (x-\hat{\mu})\in\partial\Lambda_{0}^{\hat a *}\, , \\[0.25cm]
\displaystyle   \frac{1+\gamma_{\mu}}{2}\, \psi(x) & \text{if } x\in\partial\Lambda_{0}^{\hat a} \text{ and } \exists! \; \mu\,
\big|\; (x+\hat{\mu})\in\partial\Lambda_{0}^{\hat a *}\, ,  \\[0.125cm]
   \psi(x) & \text{otherwise}\, ,
\end{cases}
\ee
and analogously for $P_{\partial\Lambda_{0}^{\hat a *}}$, i.e. with respect to $\Prj_{\partial\Lambda_{0}^{\hat a}}$ and
$\Prj_{\partial\Lambda_{0}^{\hat a *}}$  they include also the appropriate projectors on the spinor index on each face
of the boundaries. It follows that
\be
P_{\partial\Lambda_{0}} = \sum_{\hat a} P_{\partial\Lambda_{0}^{\hat a}}\; ,
\ee
and analogously for $P_{\partial\Lambda_{0}^{*}}$, $P_{\partial\Omega^{\hat a}_0}$, $P_{\partial\Omega^{\hat a *}_0}$, etc. 
The projector $P_{\partial\Pi}$ is defined as $P_{\partial\Lambda_{0}}$ but extended to all points of each hyperplane,
while $P_{\partial\Pi_1}$ is defined from 
\be
P_{\partial\Pi} = P_{\partial\Lambda_0} + P_{\partial\Pi_1}\; . 
\ee

\section{LU decomposition of a $2\,\times\, 2$ block matrix \label{app:Schur}}
A $2 \times 2$ block matrix can be decomposed as
\be\label{eq:blkdec}
M = \begin{pmatrix} A & B \\ C & D \end{pmatrix} = \begin{pmatrix} I & B D^{-1} \\ 0 & I \end{pmatrix} \begin{pmatrix} S_{A} & 0 \\ C & D \end{pmatrix},
\ee
where the Schur complement is defined as
\be\label{eq:Schur}
S_{A} = A - B D^{-1} C\; .
\ee
Its determinant can then be factorized 
\be\label{eq:detblock}
\det M = \det D\, \det \left( A - B D^{-1} C \right)\; , 
\ee
while the inverse is given by
\be\label{eq:Scmpt2}
M^{-1} = 
\left(\begin{array}{c@{~~}c@{~~}}
S_{A}^{-1} &
-S^{-1}_{A} B D^{-1}
\\[0.25cm]
- D^{-1} C S^{-1}_{A} &
\;\; D^{-1} + D^{-1} C S^{-1}_{A} B D^{-1}
\end{array}   \right)\; .
\ee
It is worth noting that $S^{-1}_{A}$ is the exact inverse of $M$ in the domain where $A$ is defined.
If $B$ and $C$ act only on subspaces identified by the projectors ${\cal P}_1$ and ${\cal P}_2$ in the first and
the second block respectively a simplification occurs, e.g. the inverse $D^{-1}$ in the second determinant on the r.h.s of
Eq.~(\ref{eq:detblock}) can be restricted to the subspace identified by ${\cal P}_2$. This in turn implies that
\be\label{eq:reduced}
\det \begin{pmatrix} A & {\cal B} \\ {\cal C} & D \end{pmatrix}
= \det A\, \det D
\det \begin{pmatrix} \id & {\cal A}^{-1} {\cal B}\\ {\cal D}^{-1} {\cal C} & \id \end{pmatrix}
\ee
where ${\cal A}^{-1} = {\cal P}_1 A^{-1} {\cal P}_1$, ${\cal B} = {\cal P}_1 B {\cal P}_2$,
${\cal C} = {\cal P}_2 C {\cal P}_1$, and ${\cal D}^{-1} = {\cal P}_2 D^{-1} {\cal P}_2$. Notice
that the dimensionality of the last matrix on the r.h.s of Eq.~(\ref{eq:reduced}) is smaller
with respect to the one of the original matrix on the l.h.s.

\bibliographystyle{JHEP}
\bibliography{mb.bib}

\end{document}